\documentclass[aps,twocolumn]{revtex4}

\usepackage{graphicx}
\usepackage{psfrag}

\begin{document} 

\title{Aharonov-Bohm effect in higher genus materials}

\author{K. Sasaki}
\affiliation{Institute for Materials Research, Tohoku University, 
Sendai 980-8577, Japan}
\author{Y. Kawazoe}
\affiliation{Institute for Materials Research, Tohoku University, 
Sendai 980-8577, Japan}
\author{R. Saito}
\affiliation{Department of Physics, Tohoku University and CREST, JST,
Sendai 980-8578, Japan}

\date{\today}

\begin{abstract}
 Flux periodicity of conducting electrons on a closed surface with genus
 two $g=2$ (double torus) are investigated theoretically.
 We examine flux periodicity of the ground-state energy and of the wave
 functions as a function of applied magnetic field. 
 A fundamental flux period of the ground-state energy is twice a
 fundamental unit of magnetic flux for uniformly applied magnetic field,
 which is shown to be valid for a simple ladder geometry and carbon
 double torus. 
 Flux periodicity of the wave functions in a double torus is complicate
 as compared with a simple torus ($g=1$), and an adiabatic addition of
 magnetic fluxes does not provide a good quantum number for the energy
 eigenstates. 
 The results are extended to higher genus materials and the implications
 of the results are discussed.
\end{abstract}

\maketitle

Geometrical structure of materials and behavior of the conducting
electrons are closely connected with each other. 
Carbon nanotube is a typical example where the electric properties are
directly related to its unique structure~\cite{SDD}.
It is known that {\it global} geometry (topology) of materials, which
consist of a closed surface, can be classified mathematically by the
number of genus ($g$).
The genus number is the number of ``holes'' of a closed orientable
surface. 
For example, as for carbon based materials,  
${\rm C}_{60}$~\cite{Kroto}, nanotubes~\cite{Iijima}, and
tori~\cite{Liu} were found in nature, and they are classified by $g=0$
or $g=1$. 
However, they are only a part of materials from the topological point of
view.
In the present letter, we examine characteristics of quantum mechanical
states of the conducting electrons in a closed surface of genus two
($g=2$, double torus). 
As electronic and magnetic properties of materials are affected by their
geometry, one may expect that, when considers different global
structures, one could find a novel phenomenon originated in the
topology.
We consider two problems associated with the topological nature of the
materials:
(1) What is a characteristic phenomenon of higher genus ($g \ge 2$)
materials that can not be expected in a lower genus ($g=0$ or $g=1$)
material? 
(2) What kind of physical quantity can be used to characterize the
energy eigenstates of the conducting electrons? 

The former question is to find a phenomenon which is closely related to
the global geometry of a material.
Aharonov-Bohm (AB) effect is an example of such a phenomenon and is one
of the most important consequence of quantum mechanics~\cite{AB}.
Ring geometries are common to investigate the AB effect where the wave
functions of electrons interfere to one another~\cite{Webb}. 
Also as for $g=2$ materials, one may expect the AB effect and the
physics should be described as a function of two independent magnetic
fields penetrating through the two holes.
There may be a general consequence of how the electrons response to the
magnetic field in higher genus materials.
The latter problem relates to taking a convenient choice of basis
vectors in the Hilbert space.
The basis vectors in ordinary materials (or bulk) are labeled by the
wave vectors.
For periodic lattice systems, one can adopt the Bloch basis vectors or
plane waves because of the lattice translational symmetry and the wave
vectors are a good quantum number.
However, as for a double torus (or higher genus materials), because of
its nontrivial topology, it seems to be difficult to define a good
quantum number.
In this letter, we examine the above two questions by analyzing the
ground-state energy, wave functions, and their periodicity as a function
of magnetic fluxes penetrating through the holes.
\begin{figure}[htbp]
 \begin{center}
  \psfrag{a}{(a)}
  \psfrag{b}{(b)}
  \includegraphics[scale=0.3]{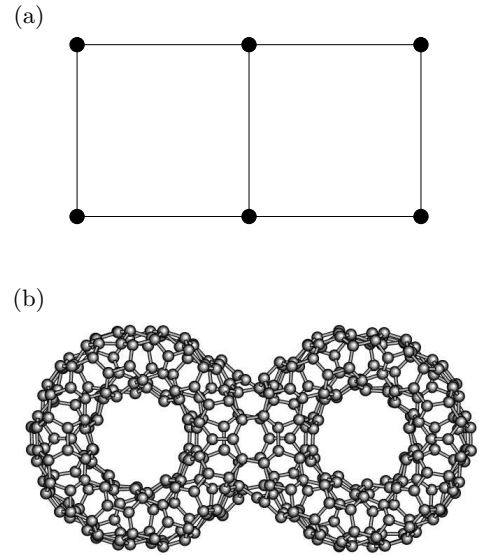}
 \end{center}
 \vspace{-1.0cm}
 \caption{(a) A simple ladder consists of six sites (Each site is
 indicated by $\bullet$).
 (b) An example of a double torus made of only carbon atoms. 
 We attach two elongated toroidal carbon nanotubes~\cite{Itoh}, both of
 which consist of 240 carbon atoms, by cutting some part of them. A
 resultant carbon double torus possesses 460 carbon atoms.
 The topological structure of (b) reduces to (a) in a limit of very thin
 tubule structure.}
 \label{fig:ng_c460}
\end{figure}

Although we consider general problems associated with the topological
nature of materials, we use two models for numerical calculations.
One is a ladder system consists of only six sites as illustrated in 
Fig.~\ref{fig:ng_c460}(a), which is thought to be a limiting shape of
$g=2$ materials. 
The other is a double torus made of carbon atoms depicted in
Fig.~\ref{fig:ng_c460}(b). 
Hereafter we will use the units: $\hbar = c = 1$.

\begin{figure}[htbp]
 \begin{center}
  \psfrag{a}{$\alpha_1$}
  \psfrag{b}{$\beta_1$}
  \psfrag{c}{$\alpha_2$}
  \psfrag{d}{$\beta_2$}
  \includegraphics[scale=0.35]{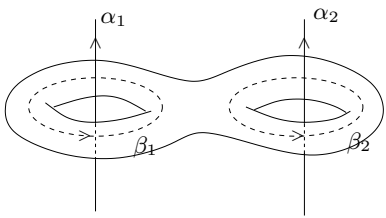}
 \end{center}
 \vspace{-0.5cm}
 \caption{A double torus with four kinds of magnetic flux. 
 The solid flux lines express a unit magnetic flux and correspond with
 gauge fields $A_1$ and $A_2$, and the dashed lines $B_1$ and $B_2$
 gauge fields respectively.}  
 \label{fig:flux_torus}
\end{figure}

First of all, we define flux lines which are necessary to analyze the AB
effect in those systems.
We define four external flux lines (or gauge fields) for a double torus
geometry: $A = \alpha_1 A_1 + \alpha_2 A_2 + \beta_1 B_1 + \beta_2
B_2$ where the vector potential $A_1(A_2)$ corresponds to a fundamental
unit of magnetic flux $\Phi_0 =2\pi/e$ ($-e$ is the electron charge)
penetrating through the left(right) hole.
$B_1$($B_2$) is the gauge field that is assigned by a magnetic flux
circling {\it inside} the surface of the left(right) ring.
The coefficients $(\alpha_1,\alpha_2,\beta_1,\beta_2)$ measure the
number of a unit flux of each component and can be a real number.
We depict these flux lines in Fig.~\ref{fig:flux_torus}.

\begin{figure}[htbp]
 \begin{center}
  \psfrag{a1}{$\alpha_2 = 0$}
  \psfrag{b1}{$\alpha_1 = \alpha_2$}
  \psfrag{a}{$\alpha_1$}
  \psfrag{c}{$\alpha_2$}
  \psfrag{x}{$\alpha_1+\alpha_2$}
  \psfrag{y}{$E(\alpha_1,\alpha_2)/V_\pi$}
  \psfrag{T1}{(a)}
  \psfrag{T2}{(b)}
  \psfrag{A}{A}
  \psfrag{B}{B}
  \includegraphics[scale=0.4]{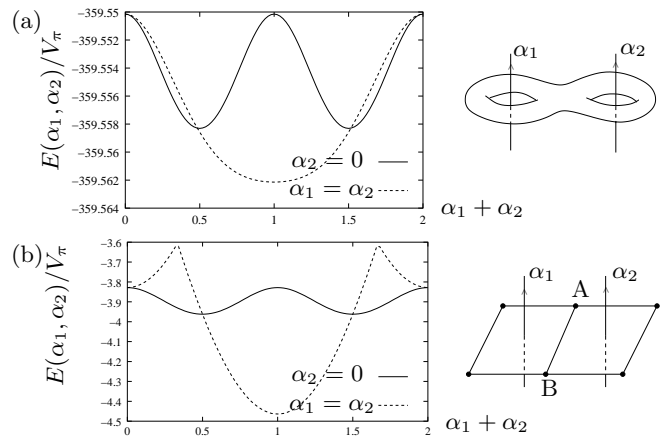}
  \end{center}
 \vspace{-0.5cm}
  \caption{(a) Numerical result of the ground-state energy of the double 
 torus made of carbon as a function of total number of magnetic flux
 ($\alpha_1+\alpha_2$).
 The solid(dashed) line indicates a flux periodicity of
 Eq.(\ref{eq:unit-flux-period})(Eq.(\ref{eq:period-doubling})).
 (b) The case of the ladder system.
 We have assumed half-filling for both cases and neglected the electron
 spin.} 
  \label{fig:vacene-dtorus}
\end{figure}

We examine the ground-state energy of the conducting electrons in
a double torus and its period as a function of applied magnetic field
expressed by $(\alpha_1,\alpha_2)$. 
The other types of flux $(\beta_1,\beta_2)$ are fixed at zero.
This is because a phenomenon peculiar to the number of genus seems to be
insensitive to the perturbation driven by $(\beta_1,\beta_2)$.
Suppose we obtain the energy eigenvalues of the Hamiltonian as
$\epsilon_i(\alpha_1,\alpha_2)$ ($i \in \{1,\cdots,N\}$ where $N$ is the
number of lattice site).
The ground state of the Hamiltonian is defined as the lowest energy
state at any external gauge field. 
The spectra of the Hamiltonian and the ground-state energy are invariant
with respect to the addition of a unit flux.
Let us define the ground-state energy as $E(\alpha_1,\alpha_2)$ and then
it has the following periodicity:
\begin{equation}
 E(\alpha_1,\alpha_2) = E(\alpha_1+1,\alpha_2) = E(\alpha_1,\alpha_2+1).
  \label{eq:unit-flux-period}
\end{equation}
The ground-state energy can be rewritten as a function of
$\alpha_1+\alpha_2$ and $\alpha_1 - \alpha_2$, so that for a uniform
magnetic field $\alpha_1 = \alpha_2 = \alpha$, it becomes a function of
only $\alpha_1+\alpha_2$, and has the following periodicity:
\begin{eqnarray}
 E(\alpha_1+\alpha_2) = E(\alpha_1+\alpha_2+2).
  \label{eq:period-doubling}
\end{eqnarray}
To check this periodicity (hereafter we call this ``period doubling''),
we have performed numerical estimation of the ground-state energy of the
carbon double torus and the ladder system
(Fig.~\ref{fig:vacene-dtorus}), assuming that the Hamiltonian of the
conducting electrons is given by the following nearest-neighbor
tight-binding Hamiltonian with an external gauge field $A$:
\begin{equation}
 {\cal H}(A) =  V_\pi \sum_{\langle i,j \rangle}
  a_j^\dagger e^{-ie\int_{r_i}^{r_j} A \cdot ds} a_i,
  \label{eq:hamiltonian}
\end{equation}
where $V_\pi$ is the hopping integral and the sum $\langle i,j \rangle$
is over pairs of nearest neighbor sites $i,j$.
The vector $r_i$ labels the vector pointing each site $i$, $a_i$ and
$a_j^\dagger$ are canonical annihilation-creation operators of the 
electrons of site $i$ and $j$ that satisfy a standard anti-commutation
relation $\{ a_i,a_j^\dagger \} = \delta_{ij}$, and $ds$ is the
differential line element.

We have numerically checked that the period doubling effect is valid in
both samples (see Fig.~\ref{fig:vacene-dtorus}).
In Fig.~\ref{fig:vacene-dtorus}(a) and (b), numerical estimation of the
ground-state energy of each sample is given.
Two curves are plotted for each sample, one is for a flux penetrating
only through the left hole(solid line) and the other is for a uniform
magnetic field(dashed line).
The periodicity of dashed lines corresponds to
Eq.(\ref{eq:period-doubling}), which is a straightforward consequence of
Eq.(\ref{eq:unit-flux-period}).

We consider a possibility that the ground-state energy has the
periodicity of a unit flux with respect of a uniform magnetic field,
that is $E(\alpha_1+\alpha_2) = E(\alpha_1+\alpha_2+1)$.
Suppose there is no hopping interaction between site A and B in the
ladder system (see the right inset of Fig.~\ref{fig:vacene-dtorus}(b)),
then the geometry of the ladder reduces to a torus ($g=1$) and the
period should become a unit flux.
Hence, the period doubling effect may be a phenomenon which reflects the
topological nature of materials.

\begin{figure}[htbp]
 \begin{center}
  \psfrag{a}{(a)}
  \psfrag{b}{(b)}
  \psfrag{x}{$\alpha_2=\alpha_3=0$}
  \psfrag{y}{$\alpha_1=\alpha_2=\alpha_3$}
  \psfrag{A}{$\alpha_1$}
  \psfrag{B}{$\alpha_2$}
  \psfrag{C}{$\alpha_3$}
  \psfrag{G}{$\alpha_1+\alpha_2+\alpha_3$}
  \includegraphics[scale=0.35]{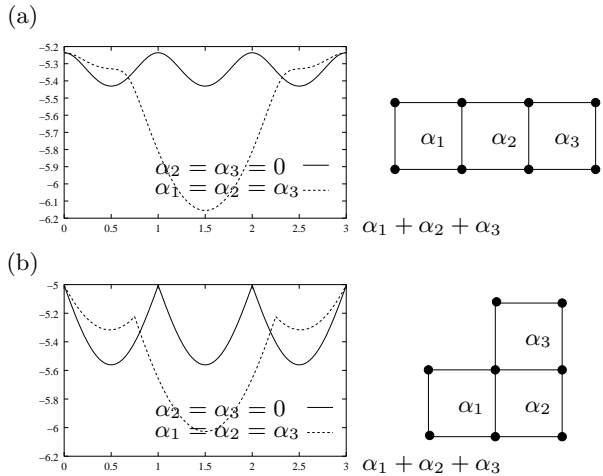}
 \end{center}
 \vspace{-0.5cm}
 \caption{(a) Numerical results of ground-state energy of ladders
 (whose geometrical configuration is shown in the right inset)
 as a function of total number of magnetic flux
 ($\alpha_1+\alpha_2+\alpha_3$).
 (b) The case of another ladder system.
 We have assumed half-filling in both calculations.}  
 \label{fig:ladder-3g}
\end{figure}

The period doubling effect can be easily extended to higher genus
materials $(g)$ under a uniform magnetic field, in which a fundamental
period of the ground-state energy can be thought of as $g$ times a flux
unit: $\Phi_{\rm unit} = g \Phi_0$. 
We have checked for this extension using two different ladder systems
shown in the right insets of Fig.~\ref{fig:ladder-3g}, which are
regarded as a limiting shape of $g=3$ materials. 
In Fig.~\ref{fig:ladder-3g}(a) and (b), numerical estimation of the
ground-state energy of each sample is given.
Two curves are plotted for each sample, one is for a flux penetrating
only through the left most hole(solid line) and the other is for a
uniform magnetic field(dashed line).
The periodicity of dashed lines is $3\Phi_0$, which is regarded as a
consequence of Eq.(\ref{eq:unit-flux-period})(including $\alpha_3 \to
\alpha_3+1$ periodicity).

We proceed to examine flux periodicity of wave functions of the
conducting electrons.
It is first noted that the flux periodicity of wave functions does not
need to be the same as that of the ground-state energy and it relates to
the detail of the geometry (or the lattice structure).
To make this point clear, let us mention the kinematics of the conducting
electrons in a torus ($g=1$) and explain periodicity of the wave
functions.
A torus can be mapped to a parallelogram with two side vectors $T_1$ and
$T_2$ for around and along the tubule axis respectively.
Because of the periodic boundary conditions along these vectors, the
wave vectors of the conducting electrons $(k)$ have to satisfy the
following constraints:
$T_1 \cdot k_1 = 2\pi n ,\ T_1 \cdot k_2 = 0,\ T_2 \cdot k_1 = 0,\ T_2
\cdot k_2 = 2\pi m$,
where we set $k = k_1 + k_2$ and both $n$ and $m$ are integers.
The wave vectors are of great use to characterize a quantum state and a
full set of them forms a complete basis in the Hilbert space.
When the Hamiltonian possesses lattice translational symmetries along
those vectors, $k$ can be used as labels for the energy eigenstates.
Here we denote the Bloch basis vectors as $|k_1,k_2 \rangle $.

We define two flux lines for a torus geometry: $A = \alpha A_1 + \beta
A_2$, where the vector $A_1$ is a gauge field corresponding to a
fundamental unit of magnetic flux penetrating through the center of a
ring and $A_2$ a unit flux circling inside the surface of a torus. 
The coefficients $\alpha$ and $\beta$ are defined as the number of a
unit flux for $A_1$ and $A_2$ respectively, and can be a real number.
Let us consider an adiabatic process in which we are adding a unit
magnetic flux by changing the number of $\alpha$ and $\beta$ gradually.  
We denote these operations as $G_1$ and $G_2$, 
$ G_1 : \alpha \to \alpha +1, G_2 : \beta \to \beta + 1$.
The difference between the Hamiltonian before and after these operations
may be thought of as the large gauge transformation~\cite{Rajaraman} and
the spectra of the Hamiltonian must be periodic in the unit of flux
quanta. 
During the adiabatic process, the eigenstates change and when we finish
adding just one unit flux, a state have to go to one of the eigenstates
in the spectra of the original Hamiltonian.
The resultant state is generally different from the original state. 
Here we define the state vectors that are obtained from an eigenstate 
by the adiabatic addition of magnetic fluxes as 
\begin{equation}
  | k_1- a eA_1, k_2-beA_2 \rangle = (G_1)^a (G_2)^b| k_1,k_2 \rangle,
\end{equation}
where $a$ and $b$ are integers.
The periodicity of the wave functions depends on the lattice structure
of a torus (or the congruent vectors, $K_i$) because $k_i$ and $k_i +
K_i$ express the same state so that the periodicity of the wave
functions can be derived as $a_i$ from the following equations: $a_i
eA_i = K_i$.
Therefore, the periodicity of the wave functions depends on its lattice
structure, and the wave vectors (or the number of magnetic fluxes) work
as good quantum numbers.
The flow of energy spectra as a function of an applied magnetic field
can be used to examine if the resultant state is different from the
original state and the periodicity of the wave functions.

\begin{figure}[htbp]
 \begin{center}
  \psfrag{a}{(a) $\alpha_- = 0$}
  \psfrag{b}{(b) $\alpha_+ = 0$}
  \psfrag{x}{$\alpha_+$}
  \psfrag{y}{$\epsilon_i(\alpha_+,\alpha_-)/V_\pi$}
  \psfrag{z}{$\alpha_-$}
  \psfrag{w}{$\epsilon_i(\alpha_+,\alpha_-)/V_\pi$}
  \includegraphics[scale=0.4]{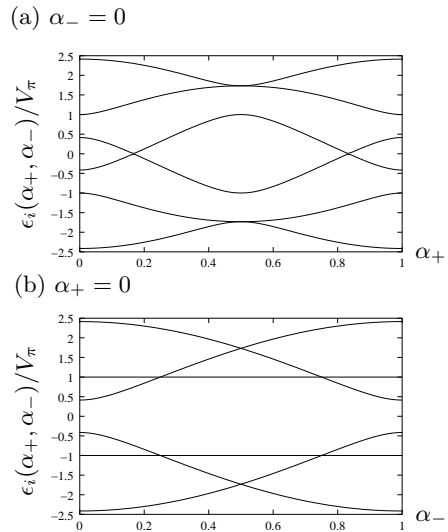}
 \end{center}
 \vspace{-0.5cm}
 \caption{Numerical results of spectra of the ladder system as a function
 of magnetic field: $2\alpha_+ = \alpha_1+\alpha_2,$
 $2\alpha_-=\alpha_1-\alpha_2$. (a) We set $\alpha_- = 0$ and vary
 $\alpha_+$, which corresponds to a uniform magnetic field. (b) We fix
 $\alpha_+ = 0$ and examine $\alpha_-$ dependence of the energy
 eigenvalue.} 
 \label{fig:spectra-ladder}
\end{figure}
\begin{figure}[htbp]
 \begin{center}
  \psfrag{x1}{$\alpha_+$}
  \psfrag{x2}{$\alpha_-$}
  \psfrag{x3}{$\beta_+$}
  \psfrag{x4}{$\beta_-$}
  \psfrag{y}{}
  \includegraphics[scale=0.33]{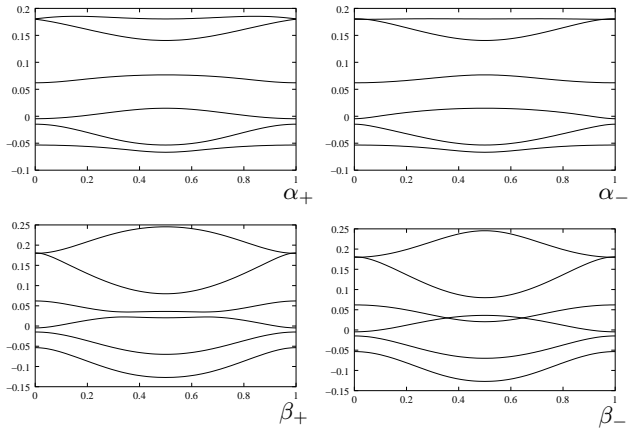}
 \end{center}
\vspace{-0.5cm}
 \caption{Numerical results of energy eigenvalues of the carbon double
 torus. Eigenvalues near the Fermi level (band center) are depicted as a
 function of magnetic field denoted in the $x$-axis of each inset, where
 we define $2\alpha_\pm = \alpha_1 \pm \alpha_2$, 
 $2\beta_\pm = \beta_1 \pm \beta_2$ (and all the other variables are
 fixed at zero).
 Each vertical axis indicates the energy eigenvalue in unit of the
 hopping integral.
 We have also checked that different flux configurations such as
 $\alpha_1 \pm \beta_1$ and $\alpha_1 \pm \beta_2$, and could not find
 the transition between different states near the Fermi level.}  
 \label{fig:spectra-dt}
\end{figure}

Let us return to the double torus.
We have examined the periodicity of the wave functions by analyzing the
spectral flow of the energy eigenvalues of the two systems shown in
Fig.~\ref{fig:ng_c460}.
We first show a numerical result of the ladder system in
Fig.~\ref{fig:spectra-ladder}.
We observe that the addition of the $\alpha_+$ magnetic flux does not
result in a connection between different states, which means that the
wave function's periodicity is the same as that of the ground-state
energy. 
On the other hand, $\alpha_-$ flux gives the transition from the
lowest(highest) energy eigenstate to a state nearest to the Fermi level.
Next we analyze the carbon double torus (see Fig.~\ref{fig:spectra-dt}).
In Fig.~\ref{fig:spectra-dt}, 
we see that any of the adiabatic processes does not provide a transition
from one energy eigenstate to a different eigenstate.
Thus, for this carbon double torus, it is difficult to assign a
quantum number by the number of magnetic flux and it suggests that 
it is better to use site basis vectors to express the energy eigenstate
near the Fermi level.

Here let us refer to some possible extensions of our results.
The ground-state energy is closely related to the persistent
currents~\cite{BIL} and the differential susceptibility, for example,
the periodicity of the ground-state energy is preserved in them.
Therefore, it could be possible to extract a kind of topological
information, e.g., whether the conducting electrons are hopping through
the line lying between two holes (hopping between site A and B in the
ladder system) or the line is cut (broken), by studying the periodicity
of persistent currents or differential susceptibility.
Because, if the hopping is not active, the period would recover a
standard periodicity of a unit flux.

Next, we comment on the periodicity of the fluxes corresponding to 
$(\beta_1,\beta_2)$.
The ground-state energy is also a function of these parameters and is
periodic in the unit of flux quanta, hence we have 
$E(\beta_1,\beta_2) = E(\beta_1+1,\beta_2) = E(\beta_1,\beta_2+1)$.
In case of $\beta_1 = \pm \beta_2$, different from the $(\alpha_1,\alpha_2)$
magnetic field, the periodicity can not be regarded as doubling because
their flux lines can be connected to form one flux line, i.e.,
$\beta_\pm=1$ corresponds to one flux line (see Fig.~\ref{fig:spectra-dt}). 

Finally, let us comment on an existing material which can be thought of
as a limiting shape of higher genus materials. 
A line of 16 GaAs/GaAlAs connected mesoscopic rings have already
manipulated and persistent currents in the rings were examined by
Rabaud et al.~\cite{Rabaud}.
In their setting, each ring is order of ${\rm \mu m}^2$, which requires
a magnetic field $B \sim 40 [{\rm gauss}]$ as a flux quanta because of
$\Phi_0 = 4 \times 10^{-7} [{\rm gauss} \cdot {\rm cm}^2]$.

In summary, we have examined periodicity of the ground-state energy and
wave functions of the conducting electrons in double torus systems under
a magnetic field. 
We have numerically checked the periodicities expressed by
Eq.(\ref{eq:unit-flux-period}) and its consequence:
Eq.(\ref{eq:period-doubling}) for the ladder and carbon double torus. 
It is expected that, for higher genus materials $(g)$, fundamental
periodicity under a uniform magnetic field is $g \Phi_0$.
It has been shown that the periodicity of the energy eigenstate near the
Fermi level is the same as that of the ground-state energy, which
indicates that it is difficult to assign a quantum number by the number
of magnetic flux.

\begin{acknowledgments}
 K. S. wishes to thank Dr. T. Tani for fruitful discussion. 
 He is supported by a fellowship of 21st century COE program of
 international center of research and education for materials of Tohoku
 university. 
 R. S. acknowledges a Grant-in-Aid (No. 13440091) from the Ministry of
 Education, Japan.
\end{acknowledgments}


\end{document}